\documentclass[twocolumn]{revtex4-1}

\usepackage{graphicx}
\usepackage{amsmath}

\begin{document}

\title{Unified rheology of vibro-fluidized dry granular media: \\ From slow dense flows to fast gas-like regimes.}

\author{Andrea Gnoli$^{1}$}
\author{Antonio Lasanta$^{1,2}$}
\author{Alessandro Sarracino$^{1}$}
\author{Andrea Puglisi$^{1,*}$} 
%\affiliation{$^1$Istituto dei Sistemi Complessi - CNR and Dipartimento di Fisica, Universit\`a di Roma Sapienza, P.le Aldo Moro 2, 00185, Rome, Italy \\
%$^*$Corresponding author}
\affiliation{$^1$Istituto dei Sistemi Complessi - CNR and Dipartimento di Fisica - Universit\`a di Roma Sapienza, P.le Aldo Moro 2, 00185, Rome, Italy \\
$^2$Departamento de F\'isica - Universidad de Extremadura, 06071 Badajoz, Spain\\ $^*$Corresponding author}

%\date{\today}

%\pacs{}

%\linenumbers

\begin{abstract}
Granular media take on great importance in industry and geophysics,
posing a severe challenge to materials
science~\cite{JNB96,andreotti13,puglio15}. Their response
properties elude known soft rheological
models~\cite{sollich97,delannay07,forterre08,henann13}, even when the
yield-stress discontinuity is blurred by
vibro-fluidization~\cite{dijksman11}.  Here we propose a broad
rheological scenario where average stress sums up a frictional
contribution, generalizing  conventional
$\mu(I)$-rheology~\cite{gdr04,cruz05,jop06}, and a kinetic collisional
term dominating at fast fluidization~\cite{volfson03}. Our conjecture
fairly describes a wide series of experiments in a vibrofluidized vane
setup~\cite{dzuy83}, whose phenomenology includes velocity weakening, shear thinning, a
discontinuous thinning transition, and gaseous shear thickening. The
employed setup gives access to dynamic fluctuations, which exhibit a
broad range of timescales~\cite{scalliet15}. In the slow dense regime
the frequency of cage-opening increases with stress and enhances, with
respect to $\mu(I)$-rheology, the decrease of viscosity. Diffusivity
is exponential in the shear stress in both thinning and thickening
regimes, with a huge growth near the transition.
\end{abstract}

\maketitle

\vspace{1cm}

\begin{figure*}[t!] \includegraphics[width=14cm]{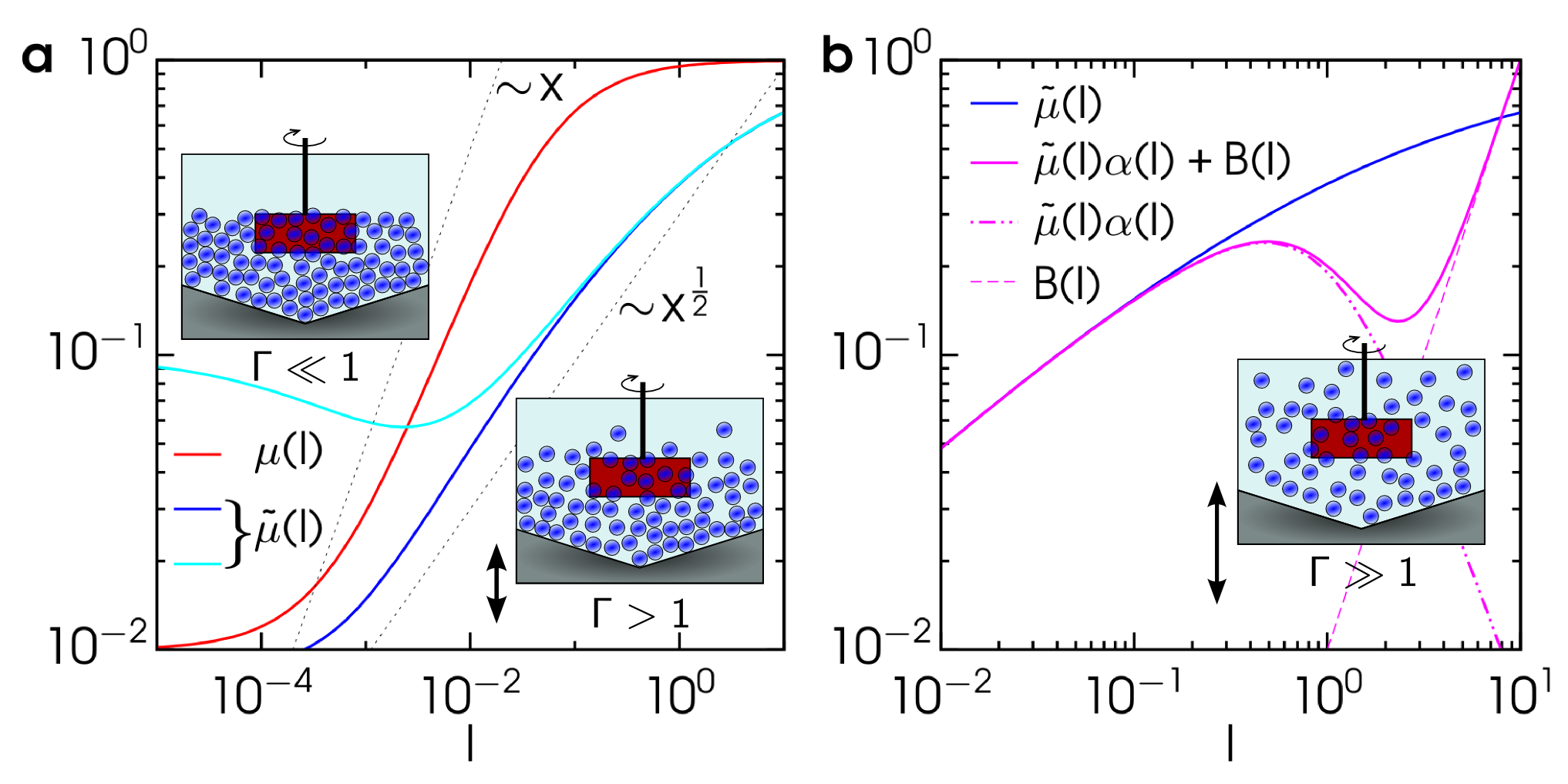}
%\raisebox{-0.05\height}{\includegraphics[width=12cm]{DATI/fig1.pdf}}
\caption{Schematic behavior of rheological functions introduced in the
  text: a) focus on low values of $I$; b) focus on larger values of
  $I$. In the two plots: $\mu(I)$ is the standard $I$-dependent
  friction coefficient, $\tilde{\mu}(I)$ is a modified version
  including the effect of activated fluidization (see
  Eq.~(\ref{model}) in the text), $\alpha(I)$ is the Bernoulli
  pressure correction and finally $B(I)$ is the Bagnold rheology
  function. Values of the constants are: $\mu_2=1$, $\mu_1=0.01$ in
  red, blue and purple curves, $\mu_1=0.1$ in cyan curve, $I_0=0.05$,
  $I_1=0.001$, $I_2=1$, $I_3=10$, $c=1$. The three drawings represent
  three characteristic regimes of fluidization: the original $\mu(I)$
  rheology describes low (or zero) fluidization, the modified
  $\tilde{\mu}(I)$ rheology includes the first effects of
  fluidization, the further modifications appearing in the full
  Eq.~\eqref{eq:unif} apply to large values of
  $\Gamma$. \label{fig:intro}}
\end{figure*}

Dry granular materials are collections of macroscopic particles,
%grains of size from hundreds
%to thousands of microns, 
interacting through frictional contact
forces. The resistance of a granular aggregate to an applied shearing
force is sensitive to many aspects of the experimental setup and may
present analogies with macroscopic frictional laws, plasticity, soft glassy
rheology and the shear thinning or thickening phenomena of
suspensions~\cite{brown10,BGP11,dullens11,kawasaki14}. Recently, consensus has been achieved on a certain
class of steady slow flows which obey the so-called
$\mu(I)$-rheology~\cite{gdr04,jop06,cruz05,forterre08}. In such a framework the shear stress $\sigma$ is
proportional to normal pressure $p$ through a friction coefficient
$\mu(I)=\sigma/p$, which slightly depends on the shear rate itself
through the adimensional ``inertial number'' $I$, according to the
following formula:
\begin{equation} \label{eq:mui}
\mu(I) = \mu_1+\frac{\mu_2-\mu_1}{1+I_0/I} = \frac{\mu_1 + \mu_2
I/I_0}{1 + I/I_0},
\end{equation} 
where $\mu_1$, $\mu_2$ and $I_0$ are constants.
The above formula (see red curve in Fig.~\ref{fig:intro}a) expresses (at constant $p$) a monotonic growth of
$\sigma$ from a minimum yield stress $\sigma_1 = \mu_1 p$ to a
saturation (frictional) stress $\sigma_2 = \mu_2 p$.  The inertial
number $I=\dot\gamma/f_m$ is the ratio between the shear rate
$\dot\gamma$ and the microscopic frequency $f_m=\sqrt{p/\rho}/d \approx
\sqrt{p/m} d^{D_s/2-1}$ ($d$ the diameter of a grain, $\rho$ its
material density, $m$ its mass, $D_s$ the space dimension).  Basically
$f_m$ is the inverse of the time needed by a grain to move by $d$ under
the acceleration given by the pressure, if starting at rest. The
validity of the $\mu(I)$ scenario has been probed in different setups
and is typically associated with a dilatancy effect in the form of a
$I$-dependent packing fraction $\phi(I)$~\cite{gdr04}. For this reason
the scenario is better appreciated in experiments where the volume is
not constrained. Note that Eq.~\eqref{eq:mui} corresponds to a
monotonic thinning-like reduction of effective viscosity
$\eta=\sigma/\dot\gamma$ which goes from $\infty$ to $0$ as the shear
rate is increased.

A more complex picture emerges in the presence of vibro-fluidization,
that is, under vertical vibration of the granular
container~\cite{danna03,dijksman11}. In applications,
vibro-fluidization is a renowned technique that enhances
homogenization and surface of contact at the solid-gas interface for
combustion chambers and chemical reactors.  A parameter that
characterizes the intensity of vibration is $\Gamma=a_{max}/g$, that
is the maximum vertical acceleration $a_{max}$ (in the case of sinusoidal
vibration) normalized by gravity acceleration $g$. Even at mild values
of $\Gamma$ ($\Gamma<1$), an internal diffusion of kinetic energy cooperates with
the applied stress and softens the discontinuities provided by
enduring contacts~\cite{dijksman11}. The result is the introduction of
a thermal-like energy scale (absent in non-fluidized granular media),
an evident reduction of the yield stress and a faster fluidization of
the material under increasing rates of deformation. Rheological
studies in a split-bottom cell under vertical vibro-fluidization
demonstrated the existence of a thinning transition~\cite{dijksman11},
whose exact nature is under scrutiny~\cite{gravish10,wulfert16},
recently ascribed to an internal distribution of microscopic stresses
and a local Herschel-Bulkley rate-stress relation~\cite{wortel16}.

A parallel line of investigation has approached the problem of dry
granular rheology by introducing the concept of partial
fluidization~\cite{volfson03,bouzid13}. In this context
there is agreement about the hybrid nature of granular internal
stress, modelled as a superposition of a frictional
contribution, sustained by enduring contacts stabilized by normal
pressure, and a kinetic contribution, where momentum is
transferred through instantaneous collisions of the fluidized
particles. The kinetic contribution is expected to be negligible in the
densest and slowest regimes, while it emerges in liquid-like flows and
finally becomes dominant in gas-like configurations. Notwithstanding
the immediacy of the concept of partial fluidization, very different
recipes and analyses have been suggested in the literature, focusing
on different aspects and setups. A relevant role in this framework is
played by models of non-local
rheology~\cite{kamrin12,henann13,bouzid13}.

Our aim, here, is to put under scrutiny a conjecture of ours for a
minimal rheological model, based upon superposition between frictional
and collisional contributions to internal stresses, that can embrace the full
spectrum of rotationally forced granular flows under
vibro-fluidization, specifically a large range of values of $I \in
[10^{-5},10]$ and $\Gamma \in [0,40]$. In general, normal stress
(pressure) $p$ depends upon the degree of fluidization, i.e. upon both
$I$ and $\Gamma$. For this reason we take as a pressure scale $p_{00}$
which is the pressure at total rest ($I=0$ and $\Gamma=0$): the
inertial number $I$ takes the same definition as above, by replacing
$p$ with $p_{00}$. Our proposal, illustrated in Fig.~\ref{fig:intro}, takes the following form for a rheological curve at constant $\Gamma$:
\begin{equation} \label{eq:unif}
\frac{\sigma}{p_{00}} = \tilde{\mu}(I)\alpha(I) + B(I),
\end{equation}
where the modified friction coefficient (blue and cyan curves in Fig.~\ref{fig:intro}a) has the form
\begin{equation}
\tilde{\mu}(I)=\frac{\mu_1 + \mu_2 I/I_0}{1 + I/I_0 + \sqrt{I/I_1}},\label{model}
\end{equation}
the Bernoulli pressure correction function $\alpha(I)$ (see dot-dashed purple curve in Fig.~\ref{fig:intro}b) is defined as
\begin{equation}
\alpha(I)=\frac{c}{1+(I/I_2)^2},
\end{equation}
and finally the Bagnold rheology function (dashed purple curve in Fig.~\ref{fig:intro}b) is simply
\begin{equation}
B(I)=(I/I_3)^2,
\end{equation}
with $\mu_1$, $\mu_2$, $c$, $I_0$, $I_1$, $I_2$ and $I_3$ model parameters.  Our proposal is
not only supported by a wide agreement with experimental data,
discussed below, but is substantiated through the following physical
arguments.

First, in contrast with the original $\mu(I)$ function, a
$\sim \sqrt{I}$ additional contribution appears at the denominator of
$\tilde{\mu}(I)$: it represents ``activated fluidization'', that is,
the enhancement of the breakage rate of enduring contacts due to the
applied stress. We note that the $I$-dependence of the friction
coefficient $\mu$ can be ascribed to the variation of the fraction
$P_s(I)$ of enduring ``solid''-like contacts, namely $\mu(I) \propto
P_s(I)$. A minimal model for $P_s(I)$ consists in neglecting memory
effects (expected to be important only at very slow shear rates) and
writing down a balance equation~\cite{MHMK13} $\partial_t P_s =
W(f \to s)(1-P_s)-W(s \to f)P_s$, whose stationary state reads
$P_s=W(f \to s)/[W(f \to s) + W(s \to f)]$, with $W(f \to s)$ and
$W(s \to f)$ the transition rates from fluid to solid state and
vice-versa, respectively. Comparison with the usual $\mu(I)$ rheology,
Eq.~\eqref{eq:mui}, suggests that $W(f \to s)$ and
$W(s \to f)$ are linear in $I$. On the contrary, the correction in the
$\tilde{\mu}(I)$, Eq.~\eqref{model}, implies that $W(s \to f)$ is
enhanced by an additional contribution $\sim \sqrt{I}$.  In
our experiment detailed below, the analysis of fluctuations provides
a transparent interpretation of such an additional term as a cage-exit frequency.
The $\sqrt{I}$-correction can also reproduce rheological behaviors of the kind shown
as the cyan curve in Fig.~\ref{fig:intro}a, i.e. cases of
velocity-weakening (an initial reduction of $\tilde{\mu}(I)$ from the
$\mu_1$ value) which appear in certain experiments at very low
vibro-fluidization. Weakening cannot be explained by the usual
$\mu(I)$ function, which is necessarily monotonic. The $\alpha(I)$
correction to pressure $p(I) \approx p_{00}/[1+(I/I_2)^2] \sim
p_{00}-\textrm{const} \cdot \dot\gamma^2$ dictates the drop in
pressure in the presence of finite fluid velocity, in analogy with
classical Bernoulli's principle.  Finally, the Bagnold rheology
function $B(I)$ provides us with the inertial contribution of
instantaneous collisions, dominating at large $I$, where one expects a
viscous contribution $\sigma \sim \gamma(I) I$ and the ``thermal''
fluctuations underlying effective viscosity are ruled by the shear
rate itself, that is $\gamma \sim I$. The Bagnold relation is usually
indicated as a case of shear thickening, even if there is no universal
consensus on whether the word ``thickening'' should be reserved for
dense suspensions, or it also applies to inertial effects arising in
diluted fluids.

Gathering all the pieces together, a general rheological curve is
obtained, an instance of which is shown as solid purple curve in
Fig.~\ref{fig:intro}b. At the transition between the solid-dominated
and the kinetic-dominated regions it is possible to observe a
non-monotonic van der Waals-like behavior of $\sigma$ which, in
stress-controlled experiments, appears as a discontinuous
thinning~\cite{dijksman11,wulfert16}. It is straightforward to verify that a
continuous change of parameters appearing in Eq.~\eqref{eq:unif}
transforms the non-monotonic crossover in a monotonic one, as seen in
the experiments. We underline that the non-monotonic crossover between
$\tilde{\mu}(I)\alpha(I)$ and $B(I)$ is clearly distinct from the
velocity-weakening effect discussed above, which belongs to the
behavior of $\tilde{\mu}(I)$ alone.

The unified rheological formula, Eq.~\eqref{eq:unif}, contains a
series of parameters which depend, among other physical aspects of the
setup, upon the intensity of vibro-fluidization $\Gamma$. We remark
that in the frictional contribution $\tilde{\mu}(I)\alpha(I)$ the
dependence on $\Gamma$ is expected to have a behavior opposite to that
in the kinetic contribution $B(I)$. Indeed, vibro-fluidization reduces
the steady fraction of enduring contacts, while increasing the thermal
agitation of flying/colliding particles. Such contrasting dependencies
neatly reflect our experimental observations, as described below.

\begin{figure*} 
\includegraphics[width=14cm]{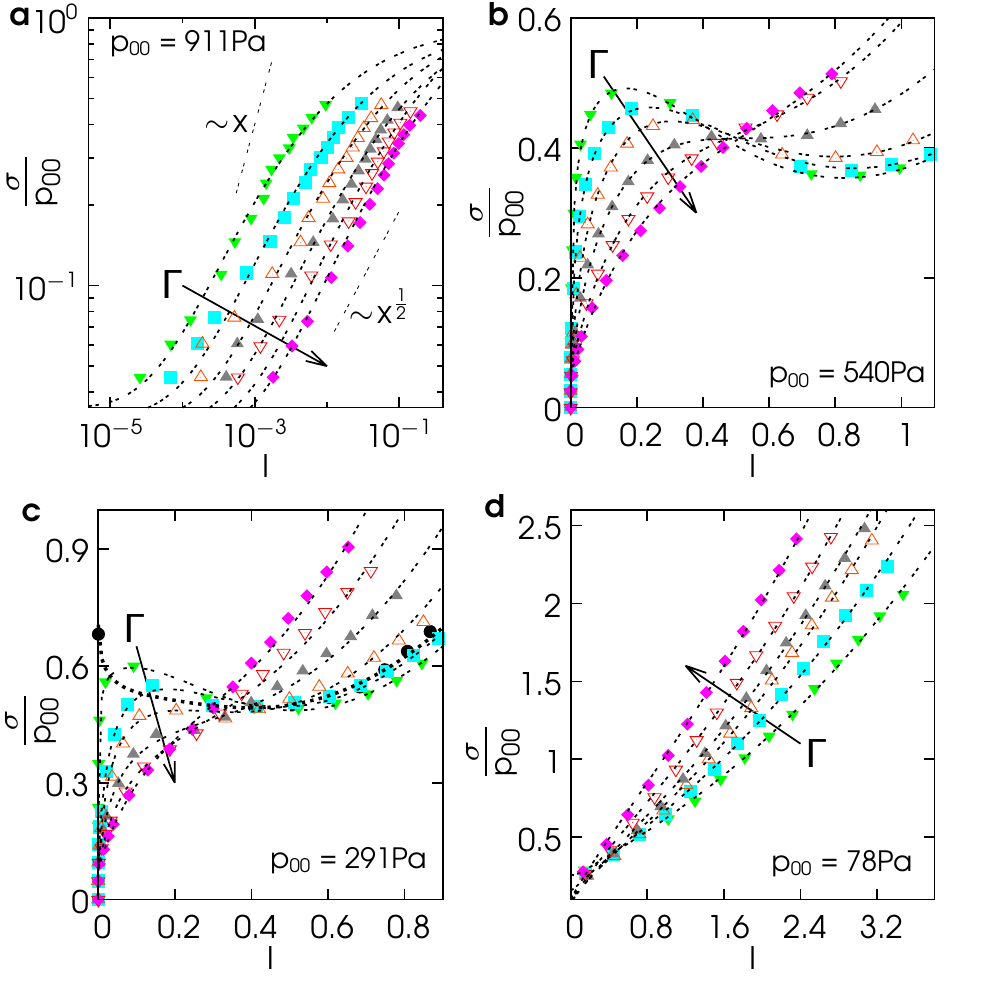}
\caption{Experimental stress-strain flow curves. Each series of data
with the same colour belongs to a value of the shaking amplitude
$\Gamma$. In frame $(a)$ the results are obtained with $N=2600$
spheres of steel, with values of $\Gamma=3.4, 6.7, 11.6, 18.3, 27.4,
38.4$ (from green to purple). In frame $(b)$ $N=1300$ spheres of
steel, with values of $\Gamma=2.4, 5, 8.9, 14.6, 22.5, 31.9$ (from
green to purple). In frame $(c)$ $N=2600$ spheres of glass, with
values of $\Gamma= 0$ (black) and $\Gamma=1.1, 8.7, 14.3, 22.1, 32,
43$ (from green to purple). Frame $(d)$ displays the results of
$N=600$ spheres of steel, with values of $\Gamma= 6.9, 8.6, 10.7,
13.2, 19.2, 26.2$ (from green to purple). Dashed lines are best fits
with Eq.~\eqref{eq:unif}. The values of the fits' parameters are given in
Table~1 in the Supplementary Information. \label{fig:flow}} 
\end{figure*}

The theoretical picture of Eq.~\eqref{eq:unif} fairly describes the
broad phenomenology observed in the experiments we carried out. These
are inspired by vane-test tools for the in-situ rheology of
soils~\cite{dzuy83,ford09}, while the granular medium
undergoes mechanical vibro-fluidization in the vertical
direction. Experiments are detailed in the Methods section.   The observed
rheological curves $\sigma$ vs $I$ explore ranges of $I$ which depend
upon $p_{00}$. The four frames in Fig.~\ref{fig:flow} show several
representative cases together with their best fits through
Eq.~\eqref{eq:unif}.

Frame $(a)$ illustrates a case at high $p_{00}$ which provides us with
a high resolution at low $\dot\gamma$, i.e. zooming in the first part
of Eq.~\eqref{eq:unif}, where the kinetic contribution is negligible
and $\alpha(I) \sim c$. The $\sim \sqrt{I}$ behavior is evident, as
well as a small but non-negligible yield stress $\mu_1>0$. At
intermediate values of $p_{00}$ (frames $(b)$ and $(c)$) the flow
curve $\sigma$ vs $I$ exhibits the crossover from the solid-dominated
regime to the collisional-dominated regime, which at low $\Gamma$ is
non-monotonic.  Increasing $\Gamma$ the parameters change
continuously, leading to a point where the curve becomes monotonic.
The pressure at rest in case $(c)$ is low enough to allow a series of
data at $\Gamma=0$ (see black circles) where a large yield stress can
be measured.  Finally, frame $(d)$ reports a low pressure situation,
where the collisional part of Eq.~\eqref{eq:unif} dominates, leading
to thickening-like behavior, that is an increasing differential
effective viscosity $\partial \sigma/\partial I$. The four frames
confirm what we argued in the above theoretical discussion: when the
stress is dominated by the solid contribution, an increase of $\Gamma$
leads to a reduction of stress, while the opposite occurs when the
kinetic contribution dominates. At a given $p_{00}$, the value of $I$
corresponding to the crossover between the two regimes does not depend
upon $\Gamma$: indeed the non-monotonic curves (case $(b)$ and $(c)$
at low $\Gamma$) cross, roughly, at a single point.

\begin{figure*} 
\includegraphics[width=14cm]{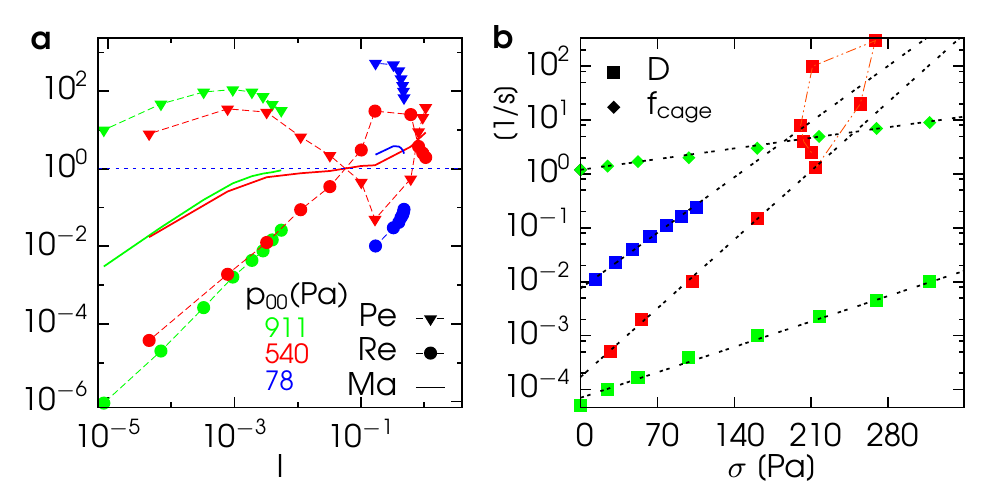}
\caption{Frame a): Peclet, Reynolds and Mach numbers, as functions of
  the inertial number $I$, in experiments at $p_{00}=911 \textrm{Pa}$
  (green symbols and lines, $2600$ spheres of steel shaken at
  $\Gamma=3.4$), at $p_{00}=540 \textrm{Pa}$ (red symbols and lines,
  $1300$ spheres of steel shaken at $\Gamma=2.4$), and at $p_{00}=78
  \textrm{Pa}$ (blue symbols and lines, $600$ spheres of steel shaken
  at $\Gamma=10.7$). Frame b): diffusivity $D$, for all three
  experiments as in frame a), and cage-exit frequency $f_{cage}$ (only
  for experiment at $p_{00}=911 \textrm{Pa}$), as function of the
  average measured stress $\sigma$. In frame $(b)$ the dashed lines
  represent exponential fits. \label{fig:fluct}}
\end{figure*}

Further support for our picture comes from the study of fluctuations,
made feasible by our vane-test experiment where the rotating blade
behaves also as a micro-rheological
probe~\cite{wang14,scalliet15,lasanta2015}. In particular we have
measured diffusivity $D = \lim_{t \gg t_0} \frac{1}{2 (t -
t_0)}|\theta(t)-\theta(t_0)|^2$ (where $\theta(t)$ is the angular
position of the blade), the frequency of relaxation of the angular
velocity $\omega(t)$ defined as $f_\omega=\langle
(\omega-\langle\omega\rangle)^2\rangle / D$, and the frequency of
typical cage exit $f_{cage}$ (which is well defined only in the slow
dense cases at high $p_{00}$~\cite{scalliet15}). The precise
definition of those quantities is given in the Methods section. The
P\'eclet number $\mathrm{Pe}=\langle \omega \rangle/D$, the Reynolds
number $\mathrm{Re} = \langle \omega \rangle/f_{\omega}$ and the Mach
number $\mathrm{Ma}=\langle \omega \rangle/\sqrt{\langle
(\omega-\langle\omega\rangle)^2\rangle}$, are shown in Fig. 3, frame
$(a)$. In both cases at high and low $p_{00}$, we find
$\mathrm{Pe} \gg 1$ and $\mathrm{Re} \ll 1$~\cite{wang16}, with the
crucial difference that $\mathrm{Ma}<1$ in the high $p_{00}$ case and
$\mathrm{Ma} > 1$ in the low $p_{00}$ case. Interestingly, an
inversion occurs - with $\mathrm{Pe}$ becoming smaller than $1$ and
$\mathrm{Re}$ larger than $1$ - at a value of $I$ corresponding to
$\mathrm{Ma}$ crossing $1$, comparable to that where the unstable
branch of $\sigma(I)$ begins. At higher $I$ the two numbers come back
to be ordered as $\mathrm{Pe} > \mathrm{Re}$, with $\mathrm{Re} >
1$. The observation of $\mathrm{Re} < 1$ in the low $p_{00}$ case with
thickening (red circles) suggests that our definition of $f_\omega$ is
not adequate in regimes of very high Mach numbers.

Green diamonds in Fig. 3b indicate $f_{cage} \sim \exp(\sigma)$ which
at low values of stress is well approximated by $f_{cage} \sim
1+\sigma$. This observation, together with the behavior
$\sigma \sim \sqrt{I}$ seen in Fig.~\ref{fig:flow}a, is compatible -
at low rates $I$ - with our interpretation of the denominator of
$\tilde{\mu}(I)$: the main responsible factor for the loosening of
solid-like contacts is the activated escape from trapping
cages~\cite{hecke15}. A further observation concerns the dependence of
$D$ on $\sigma$, again displayed in Fig. 3b: in all regimes, excluding
the dense-dilute crossover region, we observe a striking exponential
behavior $D \sim \exp(\sigma)$. This law seems universal and denotes a
wide variability of $D$ when $\sigma$ is varied keeping $\Gamma$
constant. For instance in cases near the transition a variation of
more than three decades appears. Those findings reveal an extreme
sensitivity of micro-dynamics to external disturbances which is
critical in designing industrial processes or predicting geophysical
hazards.

\section*{Methods}

\subsection{Details of the experiment}

The granular medium was made of a number $N \in [300,2600]$ of spheres
of diameter $d=4$ mm made of non-magnetic steel (mass of each sphere:
$0.267$ g), glass (mass $0.0854$ g), or
delrin\textsuperscript{\textregistered} (mass $0.0462$ g). They were
housed in a plexiglas\textsuperscript{\textregistered} cylinder with a
conical-shaped floor (diameter $90$ mm, minimum height $28,5$ mm,
maximum height $47.5$ mm) in which a plexiglas vane (height $15$ mm,
width $6$ mm, length $35$ mm) was suspended in order to be in contact
with the granular medium and not with the container~\cite{gnoli14}.  The container
was vertically vibrated by an electrodynamic shaker (LDS V450) fed
with an acceleration signal $a(t)$. In most of the experiments $a(t)$
is a white noise with a band-pass filter between $200$ Hz and $400$
Hz, while in the lowest $p_{00}$ case ($p_{00}=78$ Pa) we used a
sinusoidal signal at frequency $53$ Hz. This choice is motivated by
two empirical observations: 1) a lower number of particles (as in the
case of low $p_{00}$) requires a larger energy input to be
homogeneously fluidized and to reach the blade, and this can be
obtained by supplying energy through a sinusoidal signal
at low frequency; 2) in dense cases a sinusoidal signal induces
spurious resonances, while in diluted cases such resonances are never
observed. We have checked that performing the same experiments with
noise signal for $a(t)$ (pushing the shaker to its working limits)
gives flow curves with the same shape.  An accelerometer placed on the
container side measured $a(t)$, allowing us to define
$\Gamma=\sqrt{2\langle a^2(t)\rangle}/g$.
The vane, mounted through its rotation axis to a rotary encoder, was
also connected to a dc motor (typical working voltage $12$ V) as the
source of the driving torque. The motor was directly fed by a dc
voltage supply in the range $0$ to $7$ V. No limit was set for the
maximum current absorbed by the motor that, averaged on the duration
of the experiment, was never higher than $450$ mA. A data acquisition
system collected data for the angular position/velocity of the vane,
the effective motor voltage, the current circulating in the motor and
the root-mean-square vertical acceleration of the container. A
procedure of calibration allowed us to translate average values of
current into average values of applied torque. The same procedure
helped determining the moment of inertia of the rotating block,
$3.2\times 10^2 \textrm{g mm}^2$ (the blade with its axis and the
gears linking it to the motor). The typical experiment, at a given
$\Gamma$ and applied motor voltage, was $3600$ s long, with the
granular always ``reset'' at the beginning of each run for $30$ s at
high shaking intensity ($\Gamma=42$) and motor off. This procedure -
together with periodic replacement of used spheres - guaranteed
reproducible results at a distance of several weeks. Packing fractions
was non-homogeneous (it was larger in regions far from the borders of
the container): its value at rest was estimated to be in the range
$55\% - 70\%$, while it decreased when vibration was switched on. In
the analysis we have identified the shear rate $\dot\gamma$ with the
average of the angular velocity $\omega(t)$ of the rotating blade,
i.e. $\dot\gamma=\langle \omega(t) \rangle$, while the shear stress
$\sigma$ is proportional to the average of the applied torque $T(t)$
through the shear stress constant $\kappa$, i.e. $\sigma=\kappa
\langle T(t) \rangle$ with $\kappa=2 \pi R^2 H$ (with $R$ and $H$ the
blade half-length and height, respectively)~\cite{dzuy83}.

\subsection{Details of data analysis}

Velocity power density spectra (VPDS) are defined as $S(f)=(1/(2
t_{TOT}) |\int_0^{t_{TOT}} \omega(t) e^{i(2 \pi f) t} dt|^2$ with
$t_{TOT}$ the time-length of an experiment ($=3600$ seconds). Some
examples of $S(f)$ curves are shown in the Fig.~1 of Supplementary
Information (SI). In~\cite{scalliet15} VPDS in a similar vibro-fluidized
experimental setup, without applied torque ($\sigma=0$), has been
investigated. In the dilute or gas-like limit, e.g. low number of
spheres at high shaking, the VPDS takes a simple Lorentzian shape
$S(f)=D/[1+(2 \pi f /f_{visc})^2]$, with $D$ the asymptotic (long
time) diffusivity and $f_{visc}$ the effective viscosity due to
granular gas-vane collisions. When the number of particles (density)
is increased and/or the intensity of shaking ($\Gamma$) is reduced,
the system approaches a slow liquid regime and the VPDS develops a
wide bump (or smooth peak) with a maximum near $f \sim 20$ Hz, which
is associated to oscillations of the velocity autocorrelation induced
by liquid cages. At much smaller frequencies the VPDS reaches a
plateau whose height, $\lim_{f \to 0} S(f)$, corresponds to diffusivity
$D$: indeed the blade is not trapped in a cage forever, eventually it
manages to explore a much larger phase space and reaches normal
diffusion. From the low frequencies plateau of VPDS we have extracted
values of $D$ for Figure~\ref{fig:fluct}. We have defined the
cage-exit frequency $f_{cage}$ as the x-position, in the VPDS plot, of the
minimum separating the cage bump from the low-frequency diffusive
plateau (see filled circles in Fig.~1 of SI).

\vspace{.5cm}

{\bf Acknowledgements}

We thank Giorgio Pontuale for useful discussions and suggestions. We
also thank Rachele De Felice and Francisco Vega Reyes for reading 
the manuscript.

\vspace{.5cm}

%%%%%%%%%%%%%%%%%%%%%%%%%%%%%%%%%%%%%%%%%%%%%%%%%5
{\bf Corresponding author}

Correspondence to: Andrea Puglisi, andrea.puglisi@roma1.infn.it

\newpage

{\bf SUPPLEMENTARY INFORMATION}

\section*{S1. Parameters for the fits of rheology curves.}

In Table~\ref{tab:fits} we report the values of the parameters for the
fits of experimental data (Fig. 2 of main text) through Eq. (2) of
main text (errors are on the last digit). We notice some general trends: 

\begin{itemize}

\item $\mu_1/\mu_2$ (relative value of yield stress) decreases with $\Gamma$ (while its behavior with $p_{00}$ is not clear);

\item $I_0$ and $I_1$ increase with $\Gamma$ and decrease with $p_{00}$;

\item $I_2$ does not exhibit a strong dependence with $\Gamma$, while its dependence upon $p_{00}$ is stronger and non-monotonic;

\item $I_3$ slightly decreases with $\Gamma$ and is non-monotonic with $p_{00}$;

\end{itemize}

\begin{table*}[h!]
\begin{tabular}{|c|c|c|c|c|c|c|} \hline
  $\Gamma$ &$\mu_1/\mu_2$  &$I_0$   &$I_1$   &$c \mu_2$   &$I_2$   &$I_3$ \\ \hline \hline
\multicolumn{7}{|l|}{$p_{00}=911$ Pa ($N=2600$ spheres of steel)}\\ \hline
3.40	&0.040	&0.001	&1.2$\cdot 10^{-4}$	&0.95	&$\infty$ &$\infty$\\
6.70	&0.037	&0.003	&3.6$\cdot 10^{-4}$	&0.95	&$\infty$ &$\infty$\\
11.6	&0.033	&0.007	&7.0$\cdot 10^{-4}$	&1.03	&$\infty$ &$\infty$\\
18.3	&0.030	&0.015	&0.0015	        &1.08	&11.11 &$\infty$\\
27.4	&0.022	&0.035	&0.0030	        &1.36	&10.00 &$\infty$\\
38.4	&0.020	&0.050	&0.0032	        &1.44	&8.33 &$\infty$\\ \hline \hline
\multicolumn{7}{|l|}{$p_{00}=540$ Pa ($N=1300$ spheres of steel)}\\ \hline
2.4	&0.07	&0.0020		&4.0$\cdot 10^{-4}$	&0.65	&0.65	&2.27\\
5.0	&0.04	&0.0055	        &5.0$\cdot 10^{-4}$	&0.75	&0.69	&2.35\\
8.9	&0.04	&0.039	        &0.0080	        &0.98	&0.65	&2.21\\
14.6	&0.04	&0.10	        &0.0153	        &1.25	&0.67	&1.89\\
22.5	&0.05	&0.32	        &0.23	        &1.21	&0.71	&1.61\\
31.9	&0.04	&0.50	        &0.50	        &1.25	&0.80	&1.59\\ \hline \hline
\multicolumn{7}{|l|}{$p_{00}=291$ Pa ($N=2600$ spheres of glass)}\\ \hline
0	&0.47	&0.85		&0.38		&1.50	&0.61	&1.27 \\
1.1	&0.29	&1.1$\cdot 10^{-4}$	&2.1$\cdot 10^{-5}$ &0.66 &0.53	&1.29 \\
8.7	&0.15	&0.026		&0.795		&0.69	&0.53	&1.25 \\ 
14.3	&0.12	&0.029		&0.80		&0.62	&0.53	&1.11\\
22.1	&0.17	&0.045		&0.80		&0.55	&0.67	&1.02\\
32.0	&0.15	&0.070		&0.80		&0.55	&0.77	&0.91\\
43.0	&0.12	&0.174		&28.8		&0.67	&0.67	&0.81\\ \hline \hline
\multicolumn{7}{|l|}{$p_{00}=78$ Pa ($N=600$ spheres of steel)}\\ \hline
6.9	&0.083	&3.90	&1.44	&3.21	&16.13	&3.13\\
8.6	&0.062	&2.35	&1.36	&2.59	&9.09	&2.78\\
10.7	&0.034	&2.69	&1.71	&3.00	&10.31	&2.56\\
13.2	&0.026	&2.55	&1.75	&3.15	&7.69	&2.44\\
19.2	&0.022	&3.50	&2.66	&4.40	&7.14	&2.35\\
26.2	&0.017	&4.80	&4.18	&6.43	&7.69	&2.27\\ \hline
\end{tabular}
\caption{Table of parameters for the fits of Figure 2. \label{tab:fits}}
\end{table*}

\section*{S2. Analysis of velocity power density spectra.}

\begin{figure*} 
\includegraphics[width=18cm]{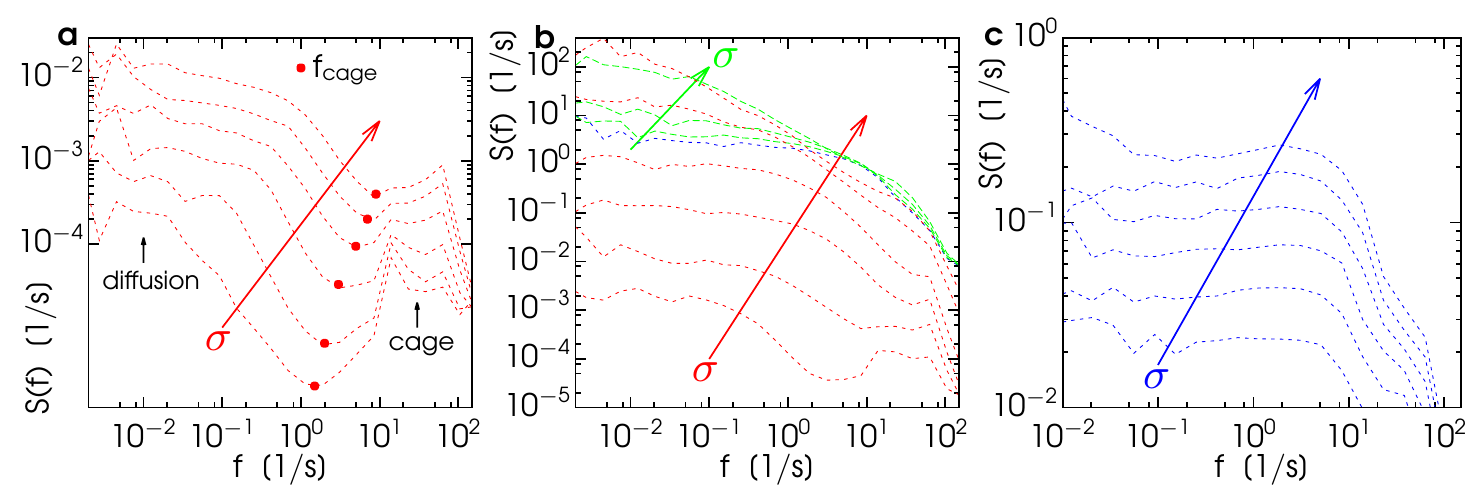}
\caption{ Power density spectra for three different values of $p_{00}$
  at mild shaking (same experiments as in Fig. 3 of main text): a) a
  case with $2600$ spheres of steel ($p_{00}=911$ Pa) at $\Gamma=3.4$,
  b) a case with $1300$ spheres of steel ($p_{00}=540$ Pa) at
  $\Gamma=2.4$, c) a case with $600$ spheres of steel ($p_{00}=78$ Pa)
  at $\Gamma=10.7$. \label{fig:vpds}}
\end{figure*}

In Figure~\ref{fig:vpds} we show some of the power density spectra of
the angular velocity time-series $\omega(t)$ measured by the blade in
$3600$ seconds-length experiments. The velocity power density spectrum is defined as
\begin{equation}
S(f)=\frac{1}{2 t_{TOT}} \left|\int_0^{t_{TOT}} \omega(t) e^{i(2 \pi f) t} dt\right|^2.
\end{equation}
In frame (a), reporting results for the high density/pressure and low
velocity experiments, we have marked with ``cage'' the bump - in the
region $10-100$ Hz, associated to fast elastic oscillations related to
trapped dynamics. In the same frame, the frequency of cage-exit
($f_{cage}$ in the main text) is the ascissa of the filled dot. At low
frequencies the height of the characteristic plateau, present in all
experiments, defines the diffusivity $D$.  In all frames the arrow
represents the order of growing $\sigma$ (shear stress or average
applied torque). In the central frame, the green curves represent the
values of $\sigma$ which decrease when $\dot\gamma$ increases (see
green triangles in Fig. 2b of main text). The blue curve corresponds
to the first point where $\sigma$ starts to grow again at large
$\dot\gamma$.

\end{document}